\documentclass[twocolumn,10pt]{article}
\usepackage[utf8]{inputenc}
\usepackage[T1]{fontenc}
\usepackage[margin=0.82in,columnsep=0.30in]{geometry}
\usepackage{amsmath,amssymb,mathrsfs}
\usepackage{comment}
\usepackage[dvipsnames]{xcolor}
\usepackage{hyperref}
\usepackage{orcidlink}
\hypersetup{colorlinks=true,linkcolor=BlueViolet,citecolor=BlueViolet,urlcolor=BlueViolet}
\usepackage{graphicx}

\def\eq#1{{Eq.~(\ref{#1})}}

\def\dd{{\rm d}}

\definecolor{oucrimsonred}{rgb}{0.6, 0.0, 0.0}
\definecolor{persianblue}{rgb}{0.11, 0.22, 0.73}
\definecolor{forestgreen}{rgb}{0.13,0.35,0.13}
\definecolor{lightgray}{rgb}{0.83, 0.83, 0.83}
 \hypersetup{colorlinks, citecolor=oucrimsonred, linkcolor=black, urlcolor=oucrimsonred}
\definecolor{cornellred}{rgb}{0.7, 0.11, 0.11}
\definecolor{navyblue}{rgb}{0.0, 0.0, 0.5}
\definecolor{amethyst}{rgb}{0.6, 0.4, 0.8}
\definecolor{yellow}{rgb}{1.0, 1.0, 0.0}
\definecolor{firebrick}{rgb}{0.7, 0.13, 0.13}
\definecolor{tangerineyellow}{rgb}{1.0, 0.8, 0.0}
\definecolor{deepfuchsia}{rgb}{0.76, 0.33, 0.76}
\definecolor{amber}{rgb}{1.0, 0.75, 0.0}
\definecolor{VioletRed4}{rgb}{0.55, 0.13, .32}
\definecolor{indiagreen}{rgb}{0.07, 0.53, 0.03}
\definecolor{VioletRed4}{rgb}{0.55, 0.13, .32}
\newcommand{\be}{\begin{equation}}
\newcommand{\ee}{\end{equation}}
\newcommand{\bea}{\begin{equation} \begin{aligned}}
\newcommand{\eea}{\end{aligned} \end{equation}}

\definecolor{oucrimsonred}{rgb}{0.6, 0.0, 0.0}
\newcommand\vertarrowbox[3][6ex]{%
  \begin{array}[t]{@{}c@{}} #2 \\
  \left\uparrow\vcenter{\hrule height #1}\right.\kern-\nulldelimiterspace\\
  \makebox[0pt]{\scriptsize#3}
  \end{array}%
}

\definecolor{verdechiaro}{rgb}{0.6,1,0.6}
\definecolor{giallochiaro}{rgb}{1,1,0.6}
\definecolor{bluscuro}{rgb}{0.15, 0.2, 0.9}
\definecolor{verdes}{rgb}{0.1, 0.5, 0.1}%
\definecolor{tangerineyellow}{rgb}{1.0, 0.8, 0.0}

\definecolor{americanrose}{rgb}{1.0, 0.01, 0.24}
\definecolor{cobalt}{rgb}{0.0, 0.28, 0.67}
\definecolor{brandeisblue}{rgb}{0.0, 0.44, 1.0}
\definecolor{mycolor}{rgb}{0.0, 0.0, 0.5}
\definecolor{oxfordblue}{rgb}{0.0, 0.13, 0.28}
\definecolor{azure}{rgb}{0.0, 0.5, 1.0}
\definecolor{turquoiseblue}{rgb}{0.0, 1.0, 0.94}

\definecolor{verdes}{rgb}{0.1, 0.5, 0.1}%
\definecolor{cornellred}{rgb}{0.7, 0.11, 0.11}

\definecolor{VioletRed4}{rgb}{0.55, 0.13, .32}

\hypersetup{
     colorlinks   = true,
     citecolor    = violet,
     urlcolor     = violet,
     linkcolor    = violet}

\def\d{{\rm d}}
\definecolor{rossocorsa}{rgb}{0.83, 0.0, 0.0}

\usepackage[normalem]{ulem}

\begin{document}

\twocolumn[
\begin{center}
{\large\bfseries On the   Modes at the Schwarzschild Black Hole Horizon}

\vspace{8pt}
Dimitrios Giataganas\,\orcidlink{0000-0003-2003-3902}$^{\,1,2*}$,
Alex Kehagias\,\orcidlink{0000-0003-4430-4914}$^{\,3,\dagger}$ \ and \
Antonio Riotto\,\orcidlink{0000-0001-6948-0856}$^{\,4,\ddagger}$

\vspace{3pt}

{\itshape\small 
$^{1}$Department of Physics, National Sun Yat-Sen University, Kaohsiung 80424, Taiwan\\
$^{2}$Physics Division, National Center for Theoretical Sciences, Taipei 10617, Taiwan\\
$^{3}$Physics Division, National Technical University of Athens,
Zografou Campus, GR-15780 Athens, Greece\\
$^{4}$D\'epartement de Physique Th\'eorique and Gravitational Wave Science Center,
Universit\'e de Gen\`eve, 24 quai E.\ Ansermet, CH-1211 Geneva, Switzerland}

\vspace{4pt}
\end{center}

\vspace{6pt}
\begin{center}
\begin{minipage}{0.9\textwidth}
\small
We discuss the origin of the  modes appearing at  the Schwarzschild black hole horizon and decaying at integer multiples of the surface gravity. We argue that no excitations generated at the horizon by the plunging of a particle into a black hole can leave it. Our argument is based on the fact that, when the particle approaches the horizon, its speed is  necessarily  boosted to ultra-relativistic values and its near-horizon gravitational field becomes a Dray-'t~Hooft shockwave, which is an exact, nonanalytic,  and  nonperturbative solution of Einstein's equations. The corresponding metric perturbations are  localized at the horizon and can never reach a distant observer.

\end{minipage}
\end{center}
\vspace{10pt}
]
\renewcommand{\thefootnote}{\fnsymbol{footnote}}
\footnotetext[1]{dimitrios.giataganas@gmail.com}
\footnotetext[2]{kehagias@central.ntua.gr}
\footnotetext[3]{antonio.riotto@unige.ch}
\renewcommand{\thefootnote}{\arabic{footnote}}
\hspace{-0.7cm}
\textbf{ \emph{Introduction.}}
When a compact object falls into a black hole, the signal reaching a distant detector
rings at the quasinormal frequencies, the genuine resonances of the geometry \cite{Berti:2025hly}. Careful
analyses of the plunge have nevertheless revealed a second family of contributions, purely
damped, with decay rates set by the surface gravity of the horizon and hence by the Hawking
temperature. The corresponding
frequencies are
\be
\omega_n=-in\kappa=-2\pi i T_{\rm H}n, \quad\quad n=0,1,\ldots,
\label{sp}
\ee
where $T_{\rm H}=\kappa/2\pi$ is the Hawking temperature and $\kappa$ the surface gravity at the horizon. If present in gravitational-wave signals, such components would probe near-horizon physics directly, since they trace the horizon's own geometry rather than the light-ring structure behind ordinary quasinormal modes.

These contributions are commonly referred to as redshift modes, or horizon
modes,  and have been identified and discussed in several formalisms~\cite{Mino:2008at,Zimmerman:2011dx,DeAmicis:2025xuh,Rosato:2026moe}.  
More recently, systematic treatments of the dynamical excitation of quasinormal
modes~\cite{DeAmicis:2025xuh,DeAmicis:2026wqd} and of the direct-wave part of the
ringdown signal~\cite{Oshita:2025qmn} have revived the question of whether this
ladder constitutes an independent, detectable set of degrees of freedom.

A sharp global statement was recently given in Ref.~\cite{Kuntz:2026xep}, which showed that, based on causality arguments,  the source-convolved waveform of a particle plunging into a Schwarzschild
black hole contains no isolated terms proportional to $e^{-n\kappa u}$, with $u$
the retarded time of the distant observer. This conclusion has been subsequently debated in Ref. \cite{Ma:2026hcb}. 

The goal of this note is to elucidate the origin of the tower of modes present at the horizon and to argue that they may not leave any imprint on the signal reaching a distant observer; as such, they are not observable. The reason, we will argue, lies in the gravitational field of
ultra-relativistic sources. Since the classic boost of the Schwarzschild
geometry performed by Aichelburg and Sexl~\cite{Aichelburg:1970dh}, it
has been known that the field of a particle taken to the speed of light
at fixed momentum collapses into an impulsive plane-fronted shockwave,
supported on a single null surface. Dray and
't~Hooft~\cite{Dray:1984ha} showed that an exact, nonperturbative
solution of Einstein's equations of the same impulsive type exists for a
(nearly) massless particle propagating along the horizon of a Schwarzschild black
hole. According to Dray and
't~Hooft, this horizon shock may be interpreted as the late-time exterior near-horizon description of a sufficiently light infalling particle. Schwarzschild time translations act as boosts in the Kruskal plane, so that the particle is seen by a stationary exterior observer with an increasingly large boost as it approaches the horizon. We shall supplement this physical interpretation with a controlled distributional construction in which a family of timelike point particle sources approaches a null source supported on the future horizon. This family provides a precise realization of the horizon source, although it should not be identified with the invariant late-time evolution of one fixed timelike worldline. We then show that the resulting Dray–’t Hooft junction carries no propagating Regge–Wheeler–Zerilli (RWZ) master degree of freedom. 

It is interesting to notice that the  situation is entirely analogous to that of the 
quasinormal mode spectrum  of extremal Kerr black holes, which contains frequencies with vanishing imaginary part and hence zero damping: such real-axis resonances are unphysical as natural black hole oscillations and are always associated with scattering modes \cite{Richartz:2017qep}.

We proceed by first commenting on   the origin of the horizon modes and linking it to the (implicitly assumed, but not proven) analyticity of the waveform in the appropriate Kruskal coordinates and subsequently  by  showing  how the exact, nonperturbative solution of the gravitational field of an ultra-relativistic particle plunging into a black hole can be provided by the  Dray-'t~Hooft shockwave. This second step demonstrates  the lack of analyticity and why the modes are not  emitted by the horizon. 

\medskip\noindent
\hspace{-0.2cm}
\textbf{\emph{The origin of the horizon tower.}} Let us  consider   a Schwarzschild black hole spacetime geometry in Kruskal-Szekeres coordinates (setting $G_N$ and $c$ to unity),
\begin{eqnarray}
\d s^2&=&2 A(U,V) \d U\d V+g(U,V) \d\Omega^2,\nonumber\\
A(U,V)&=&-\frac{16m^3}{r}e^{-r/2m},\nonumber\\
g(U,V)&=&r^2,~~~~ UV=-\left(\frac{r}{2m}-1\right)e^{r/2m},
\label{ul}
\end{eqnarray}
where $m$ is the mass of the black hole and $\d\Omega^2$ is  the metric of the two-dimensional sphere. 
The null coordinates $U,V$ are related to the Schwarzschild  coordinates through the relations
\begin{eqnarray}\label{eq:uv}
U=- e^{-\kappa u},\qquad V= e^{\kappa v},
\end{eqnarray}
with 
\begin{eqnarray}
u=t-r_*, \qquad v=t+r_*
\end{eqnarray}
where $r_*=r+2m \log(r/2m-1)$.  The black hole exterior is described by the region ($U<0,V>0$), whereas the black hole interior  is the region ($U>0,V>0$). Therefore,  the future horizon $\mathscr{H}_+$ is at $U=0$. Similarly, the region $(U>0,V<0$) is the other exterior ("the other universe") whereas $(U<0,V<0)$ describes a white hole interior. Again,  $V=0$ is the past horizon $\mathscr{H}_-$.

 Now, beyond the quasinormal modes, the modes generated at the horizon have been identified as  degrees of freedom with a wavefunction of the type
\begin{equation}
\psi(u)\supset\sum_{n\geq1}\mathcal A_n e^{-n\kappa u},
\label{eq:tower1}
\end{equation}
where $\mathcal A_n$ denotes the coefficient in the exterior time-domain expansion.  

Our first point is that  the local origin of these contributions is a simple change of frame.
Near the future horizon the Schwarzschild retarded coordinate $u$
diverges and does not extend regularly across the horizon, whereas the
outgoing Kruskal coordinate $U$ crosses $U=0$ smoothly. Assuming that the relevant outgoing field
is analytic in $U$,  it admits the convergent expansion
\begin{equation}
\psi(U)=a_0+\sum_{n\geq1}a_n U^n,
\label{eq:directtower1}
\end{equation}
where $a_n$ are its Kruskal Taylor coefficients.  Using Eq.~\eqref{eq:uv}, the Kruskal expansion becomes
\begin{equation}
\psi(u)=a_0+\sum_{n\geq1}(-1)^n a_n e^{-n\kappa u},
\end{equation}
when expressed in the Schwarzschild retarded coordinate $u$.
Consequently, the exterior redshift amplitudes Eq.~\eqref{eq:tower1} and the Kruskal Taylor
coefficients  Eq.~\eqref{eq:directtower1} are related by $\mathcal A_n=(-1)^n a_n$. The purely damped sequence is therefore nothing more mysterious than the Taylor expansion of a horizon-analytic Kruskal field, re-read in the retarded coordinate $u$.

The terms ``redshift modes'' and ``horizon modes'' refer therefore to nothing other than  the local sequence generated by the near-horizon Taylor expansion. Related contributions have also been discussed under the broader heading of direct waves~\cite{Oshita:2025qmn}. The complete direct-wave
signal, however, need not be exhausted by the local sequence of purely
damped frequencies $\omega_n=-in\kappa$, which we refer to as the
Matsubara ladder since the frequency spacing is fixed by the Hawking temperature.

The point worth stressing is that the appearance of the horizon mode tower is a consequence of analyticity, which we assumed. We will show later on that this assumption does not hold and that, as a consequence,   the corresponding horizon modes are not realized as observable excitations. In other words, they do not leave the horizon and therefore cannot be detected by a distant observer. Before turning to this point, however, let us first discuss the properties of such modes. 

\medskip\noindent
\hspace{-0.2cm}\textbf{ \emph{The ladder as a Mellin spectrum.}} 
Under a Killing-time translation $t\mapsto t+\tau$, the Kruskal coordinates transform as
\begin{equation}
U\mapsto e^{-\kappa\tau}U\,,
\qquad
V\mapsto e^{\kappa\tau}V\, ,
\end{equation}
so that the exterior time translation is   a boost of the $(U,V)$ plane.
For the outgoing sector, where the field depends on $U$, the
corresponding generator is
\begin{equation}
L_U=-\kappa U\partial_U\,.
\label{eq:generator}
\end{equation}
The usual frequency modes $F_\omega(U) \propto e^{-i\omega u}$  satisfy
\begin{equation}
L_U F_\omega=-i\omega F_\omega\,,
\label{eq:generatoreig}
\end{equation}
and they turn out to be 
\begin{equation}
F_\omega(U)=(-U)^{i\omega/\kappa}=e^{-i\omega u}\,, 
\label{eq:boostmode}
\end{equation}
so that exterior frequency modes are powers of the Kruskal
coordinate.

Two classes of functions must now be kept distinct. On
the exterior region $U<0$, the unitary boost modes
$(-U)^{i\omega/\kappa}$ are labelled by a continuous real frequency
$\omega$, which is the basis in which an exterior observer decomposes
the signal. Near the horizon, however, suppose that the outgoing field is
analytic in $U$, so that it admits the convergent expansion
\begin{equation}
F(U)=\sum_{n=0}^{\infty}a_n U^n
\label{eq:taylor}
\end{equation}
in a neighborhood of $U=0$. Each monomial is an eigenfunction of the
same generator \eqref{eq:generator},
\begin{equation}
L_U U^n=-n\kappa U^n=-i\omega_n U^n\,,
\end{equation}
with the complex frequency
\begin{equation}
\omega_n=-in\kappa\,,\qquad n=0,1,2,\ldots,
\label{eq:weights}
\end{equation}
or, equivalently,
\begin{equation}
U^n=(-1)^n e^{-n\kappa u}\,.
\end{equation}
The term $n=0$ is the constant, zero-frequency component, whereas the
purely damped horizon ladder discussed below begins at $n=1$.
Therefore, expressing the Taylor expansion of a horizon-analytic
Kruskal field in terms of the retarded coordinate $u$ gives
\begin{equation}\label{expansion1}
F(u)=a_0+\sum_{n\geq1}(-1)^n a_n e^{-n\kappa u},
\end{equation}
where the terms with $n\geq1$ form the purely damped horizon ladder,
with frequencies $\omega_n=-in\kappa.$

The same structure is encoded by a local Mellin transform, evaluated along the imaginary axis of the Mellin variable which isolates the contribution from the horizon endpoint.  Setting
$z=-U>0$, we define
\begin{equation}
\widetilde F_{\rm H}(\omega)=\int_0^{\epsilon_0}\frac{dz}{z}\,z^{-i\omega/\kappa}F(-z),
\end{equation}
where $\epsilon_0$ lies within the region in which the near-horizon
power expansion is valid. Changing the cutoff within the domain of validity of the near-horizon expansion adds a function entire in $\omega$ and therefore leaves these pole locations and residues unchanged.  

Substituting the near-horizon expansion from Eq.~\eqref{eq:taylor}, 
and integrating in the domain of convergence gives
\begin{equation}
\widetilde F_{\rm H}(\omega)=\sum_{n=0}^{\infty}(-1)^n a_n
\frac{\epsilon_0^{\,n-i\omega/\kappa}}{n-i\omega/\kappa}.
\label{eq:mellinseries}
\end{equation}
This expression provides a meromorphic continuation with 
simple poles at
\begin{equation}
\omega=\omega_n=-in\kappa,
\end{equation}
whose residues are
\begin{equation}
\operatorname*{Res}_{\omega=-in\kappa}\widetilde F_{\rm H}(\omega)=i\kappa(-1)^n a_n.
\label{eq:mellinres}
\end{equation}
This is the standard Mellin correspondence between a power
expansion near an integration endpoint and the poles of its Mellin
transform~\cite{Flajolet1995MellinTA}. The residue at the $n$th frequency
is therefore proportional to the $n$th coefficient in the analytic
near-horizon expansion.

The horizon spectrum is therefore the discrete Mellin-pole spectrum
associated with the action of near-horizon boosts on regular
Kruskal fields. Its spacing is fixed by the surface gravity and by
regularity at the horizon,
\begin{equation}
\omega_{n+1}-\omega_n=-i\kappa\,,
\end{equation}
rather than by the detailed form of the black hole potential. After
Euclidean continuation, the boost angle has period $2\pi$, and the
same frequencies may be written as the spectrum (\ref{sp}), 
which explains their Matsubara form.

This origin should not be confused with that of quasinormal modes.
Quasinormal frequencies are determined by the global scattering
problem, after imposing ingoing boundary conditions at the horizon
and outgoing boundary conditions at infinity. The horizon frequencies, 
by contrast, follow locally from the relation
between Killing time and the regular Kruskal coordinate. They are
dilation weights of the near-horizon expansion, not by themselves
dynamical resonances of the black hole.

We are now ready to take a further logical step based on a simple point. Let us consider a particle (or another much lighter black hole)  hitting a Schwarzschild black hole. Approaching its horizon, the particle's speed will be boosted and become ultra-relativistic. The process has a dramatic effect on the spacetime geometry. 
The gravitational field of the particle becomes infinitely Lorentz-contracted along its direction of motion. It no longer looks like the weak spread-out field of a static object. Instead, it gets ``squashed" into an infinitely thin, plane-fronted sheet of energy, a shockwave \cite{Aichelburg:1970dh}. 
 
 When approaching the black hole and coming  closer to the horizon, such a  plane  becomes gradually more spherical, as seen by an outside observer. The full spacetime geometry is described by the 
 Dray-'t Hooft solution \cite{Dray:1984ha}, which is the exact, nonperturbative result of considering  a Schwarzschild black hole in the presence of the shockwave at its horizon. 
 
 In other words, the process of a particle hitting a black hole should  be described, during its last stages, when the particle approaches the black hole horizon, as a corresponding shockwave  superimposed on the  Schwarzschild black hole spacetime. This is what we show now before we present our final argument.

\medskip\noindent
\hspace{-0.3cm}
\textbf{ \emph{From plunging to  shockwave.}}
The analytic horizon ladder describes fields that are regular in
Kruskal coordinates.  We now contrast this analytic sector with a
singular limit in which a family of timelike point particle sources
approaches the  black hole horizon.  This limit
must not be confused with the late time evolution of one fixed
timelike geodesic.  A fixed geodesic remains timelike and crosses the
future horizon transversely in regular coordinates.

For clarity we consider radial motion and use units in which $2m=1$.
We denote the rest mass of the particle by $m_{\rm P}$ and its
conserved specific energy by ${\cal E}$.  The radial geodesic
equations are
\begin{equation}
\frac{\dd r}{\dd\tau}=-\sqrt{{\cal E}^{2}-A(r)},\quad \frac{\dd t}{\dd\tau}=\frac{{\cal E}}{A(r)},
\label{eq:dr-dt}
\end{equation}
where
$A(r)=1-1/r,$ and the minus sign in the first equation in \eqref{eq:dr-dt} selects the ingoing branch.
We now write the energy $\mathcal{E}$ as $\epsilon$ and take the limit $\epsilon\to 0$.  The radial equation requires then $A(r)\leq\epsilon^{2}$, and therefore, the worldline is confined to an increasingly small
neighborhood of the horizon.  This family does not describe
particles falling from infinity.  It provides a controlled limit in
which a timelike trajectory becomes tangent to the future horizon.

Along the radial trajectory one finds
\begin{eqnarray}\label{eq:trajectory}
\frac{\dd v}{\dd\tau}=\frac{{\cal E}-\sqrt{{\cal E}^{2}-A}}{A}=\frac{1}{{\cal E}+\sqrt{{\cal E}^{2}-A}}, 
\end{eqnarray}
so that, at the future horizon we have
\begin{equation}
\left.\frac{\dd v}{\dd\tau}\right|_{\mathscr H_+}=\frac{1}{2\epsilon}.
\end{equation}
Similarly, using that the surface gravity is
$\kappa=\frac12$ and therefore 
$V=e^{v/2},$
we obtain
\begin{equation}
\left.
\frac{\dd V}{\dd\tau}\right|_{\mathscr H_+}=\frac{V_H}{4\epsilon},
\label{Vvelocity}
\end{equation}
where $V_H$ denotes the value of $V$ at the horizon crossing. 
The exact relation between the Kruskal coordinates and the radial 
coordinate is $-UV=e^r(r-1).$  Differentiating this relation along the trajectory and evaluating it
at $U=0$ and $r=1$ we get
\begin{equation}
-V_H\left.\frac{\dd U}{\dd\tau}\right|_{\mathscr H_+}
=e\left.\frac{\dd r}{\dd\tau}\right|_{\mathscr H_+}.
\end{equation}
At the horizon we get from \eq{eq:dr-dt},
\begin{equation}
\left.\frac{\dd r}{\dd\tau}\right|_{\mathscr H_+}=-\epsilon
\end{equation}
and therefore
\begin{equation}
\left.\frac{\dd U}{\dd\tau}\right|_{\mathscr H_+}=\frac{e\epsilon}{V_H},
\label{Uvelocity}
\end{equation}
so that Eqs. ~\eqref{Vvelocity} and \eqref{Uvelocity} imply
\begin{equation}
\left.
\frac{\dd U}{\dd V}\right|_{\mathscr H_+}=\frac{4e\epsilon^{2}}{V_H^{2}}.
\label{horizonslope}
\end{equation}
We see that the slope vanishes as $\epsilon^{2}$, and therefore, the trajectories become
tangent to $U=0$, with  direction along increasing $V$. We verify that this limit holds on a finite interval of
the horizon and not only at the crossing point.  We can show that it has the near horizon velocity
\begin{equation}
u_{\epsilon}^{V}=\frac{\dd V}{\dd\tau}=\frac{V_H}{4\epsilon}+{\cal O}(\epsilon),
\label{uniformVvelocity}
\end{equation}
in agreement with \eqref{Vvelocity}. Therefore, along the whole nearby horizon segment, $u_\epsilon^V\sim\epsilon^{-1}$ and the trajectory moves very rapidly along the $V$-direction. 
We also obtain
\begin{equation}
U_{\epsilon}(V)=\frac{4e\epsilon^{2}}{V_H^{2}}\left(V-V_H\right)+{\cal O}(\epsilon^{4}),
\label{worldlinelimit}
\end{equation}
and thus, $U_{\epsilon}(V)$ tends to zero uniformly on every compact interval of $V$ within the near horizon branch. This shows that the higher derivatives along the near-horizon branch are subleading in $\epsilon$, so the leading behavior is entirely determined by the horizon slope \eqref{horizonslope}.

We now proceed to determine the  stress tensor in the $\epsilon\to 0 $ limit.  The stress tensor of a
point particle is
\begin{equation}
T_{\epsilon}^{\mu\nu}(x)=m_{\rm P}
\int\dd\tau\,\frac{u_{\epsilon}^{\mu}u_{\epsilon}^{\nu}}{\sqrt{-g}}\delta^{(4)}\bigl(x-x_{\rm P}(\tau)\bigr).
\label{timelikeT}
\end{equation}
Since $V$ is monotonic along the future directed trajectory, as shown in \eqref{uniformVvelocity} we may
use it as the worldline parameter.  Eq.~\eqref{timelikeT} then
becomes
\begin{equation}
T_{\epsilon}^{\mu\nu}=\frac{m_{\rm P}}{\sqrt{-g}}
\frac{u_{\epsilon}^{\mu}u_{\epsilon}^{\nu}}{u_{\epsilon}^{V}}
\delta\bigl(U-U_{\epsilon}(V)\bigr)\delta^{(2)}(\Omega-\Omega_0),
\label{TparamV}
\end{equation}
and the component directed along the limiting null trajectory is
\begin{equation}
T_{\epsilon}^{VV}=\frac{m_{\rm P}u_{\epsilon}^{V}}{\sqrt{-g}}
\delta\bigl(U-U_{\epsilon}(V)\bigr)\delta^{(2)}(\Omega-\Omega_0).
\label{eq:TVV}
\end{equation}
Eq.~\eqref{uniformVvelocity}  shows that $u_{\epsilon}^{V}$ diverges as $\epsilon^{-1}$.  To obtain a finite null-source limit, we choose the rest mass according to
\begin{equation}
m_{\rm P}=\frac{4p_H}{V_H}\epsilon
\label{massscaling}
\end{equation}
where $p_H$ is held fixed.  We then have from  Eq.~\eqref{Uvelocity}   and Eq.~\eqref{uniformVvelocity}
\begin{equation}
m_{\rm P}u_{\epsilon}^{V}=p_H+{\cal O}(\epsilon^{2}),\quad m_{\rm P}u_{\epsilon}^{U}={\cal O}(\epsilon^{2}).
\end{equation}
All components that contain a transverse factor $u_{\epsilon}^{U}$ therefore vanish in the limit $\epsilon \to 0$, and only the component $T^{VV}$ survives. The latter turns out to be at the horizon 
\begin{equation}
T^{VV}=\frac{p_H}{\sqrt{-g_H}}\,\delta(U) \delta^{(2)}(\Omega-\Omega_0),
\label{contravariantshock}
\end{equation}
or after lowering the indices, the only nonzero null component is
\begin{equation}
T_{UU}^{\rm H}={\cal N}_H p_H \delta(U)\delta^{(2)}(\Omega-\Omega_0)
\label{shockstress}
\end{equation}
where ${\cal N}_H$ depends on the normalization of the Kruskal
coordinates.  This is locally the stress tensor of a massless
particle propagating along the future horizon.  The standard Dray
and 't~Hooft source is obtained by extending this null source along
the horizon generator.

We have therefore obtained the horizon null source as a controlled
limit of timelike point particle stress tensors.  The limit is
${\cal E}\to0$, together with $m_{\rm P}\to0$
so that $m_{\rm P}/{\cal E}$ is fixed.

\begin{figure}[htbp]
\centering
\includegraphics[width=\linewidth]{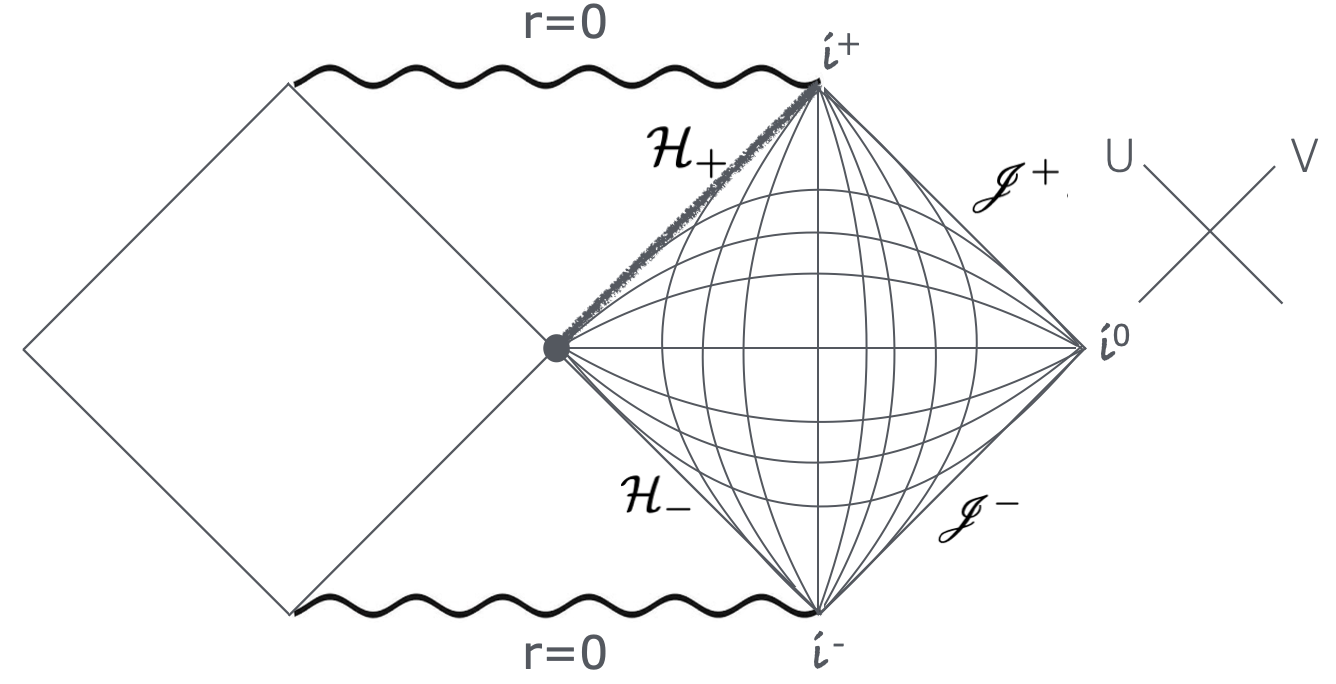}
\caption{The Penrose diagram for a shockwave supported on the
future horizon $\mathcal H_+$ of a Schwarzschild black hole.}
\label{figpendiag}
\end{figure}

\medskip\noindent
\hspace{-0.2cm}\textbf{ \emph{The nonperturbative Dray-'t~Hooft shockwave and why horizon modes are not observable.}}
Following Dray and 't Hooft \cite{Dray:1984ha}, the late-time exterior description of a light particle approaching the Schwarzschild horizon may be interpreted as a horizon shock. Since Schwarzschild time translations act as Lorentz boosts in the Kruskal plane, the particle appears increasingly boosted to an observer at fixed radius, while its gravitational field becomes concentrated on the future horizon. In this sense, the Dray–'t Hooft solution is the Schwarzschild counterpart of the Aichelburg–Sexl shock, with its wavefront adapted to the spherical geometry of the horizon. Then, the limiting source in Eq.~\eqref{shockstress} along the horizon generator gives the source of the Dray and 't~Hooft geometry~\cite{Dray:1984ha}, with the corresponding Penrose diagram  shown in
Fig.~\ref{figpendiag}.

On the side $U<0$ the geometry is the unshifted Schwarzschild
geometry
\begin{equation}
\dd s^2=2A(U,V)\dd U\dd V+g(U,V)\dd\Omega^2.
\label{unshiftedshock}
\end{equation}
Across $U=0$ the two sides are related by the null shift
\begin{equation}
V_+=V_-+f(\Omega),
\label{shockjunction}
\end{equation}
since the source is localized at a particular point of the horizon two-sphere. 
The angular profile obeys
\begin{equation}
\left(-\Delta_{S^2}+1\right)f(\Omega)=8\pi {\cal N}_H p_H \delta^{(2)}(\Omega),
\label{shockprofileequation}
\end{equation}
where the source on the right-hand side is the localized null stress tensor of Eq. \eqref{shockstress}. We can expand $f(\Omega)=\sum_{\ell,m} f_{\ell m}Y_{\ell m}(\Omega)$ in spherical harmonics, and by using the harmonic expansion of the delta function on the sphere for a particle located at the north pole we obtain
\begin{equation}
f(\Omega)=4\sqrt{\pi}\,\mathcal{N}_Hp_H\sum_{\ell\geq0}\frac{\sqrt{2\ell+1}}{\ell(\ell+1)+1}
Y_{\ell0}(\theta).
\label{shockprofile}
\end{equation}
We write the discontinuous coordinate transformation \eqref{shockjunction} as $V\longrightarrow
V+\Theta(U)f(\Omega)$ 
and the metric can be written in the distributional form
\begin{equation}
\dd s^2=2A(U,V)\dd U\left[\dd V-f(\Omega)\delta(U)\dd U\right]+g(U,V)\dd\Omega^2.
\label{ub}
\end{equation}
This is an exact nonlinear geometry.  It is smooth on either side
of $U=0$ and its physical content is the finite junction in
Eq.~\eqref{shockjunction}.

The $\delta(U)$ term in Eq. \eqref{ub} makes the metric distributional and nonanalytic across $U=0$, so it does not belong to the smooth Taylor–Mellin sector discussed above. This fact alone does not establish the absence of radiation. To determine whether the shock carries a propagating perturbative degree of freedom, we now extract the part linear in the shock profile and evaluate the corresponding gauge-invariant RWZ master function.
Expanding the metric~(\ref{ub}), we can distinguish the background
Schwarzschild geometry from the impulsive perturbation
\begin{align}
\dd s^2 {}&= \underbrace{-\frac{32 m^3}{r} e^{-r/2m} \dd U\,\dd V + r^2 \dd\Omega^2}_{\text{background}}
\nonumber\\
&+ \underbrace{\frac{32 m^3}{r} e^{-r/2m} f(\theta)\, \delta(U)\, \dd U^2}_{\text{perturbation}},
\end{align}
where we have reintroduced the mass parameter. 
Recalling that the  retarded Kruskal coordinate $U$ is defined as
\begin{align}
U &=-\sqrt{\frac{r}{2m} - 1}\; e^{r/4m}\, e^{-t/4m},
\end{align}
the perturbed metric becomes
\begin{align}
\dd s^2_{\text{pert}} = \Phi \left( \dd t - \frac{\dd r}{1 - 2m/r} \right)^2,
\end{align}
where
\begin{align}
\Phi ={}& \frac{2m}{r} e^{-r/2m} f(\theta)\, U^2\, \delta(U).
\end{align}
Now, the standard even-parity perturbation in the Regge-Wheeler gauge is
\begin{align}
h_{\mu\nu} \dd x^\mu \dd x^\nu ={}& -\left(1-\frac{2m}{r}\right) H_0\, \dd t^2 + 2 H_1\, \dd t\, \dd r
\nonumber\\
&+ \frac{H_2}{1-\frac{2m}{r}}\, \dd r^2 + r^2 K\, \dd\Omega^2.
\label{h}
\end{align}
Expanding in multipoles and around the horizon, with $H_i(t,r,\Omega)=\sum_{\ell,m}
H^{(i)}_{\ell m}(t,r)Y_{\ell m}(\Omega)$,  we find that the even perturbations due to this shockwave in the Regge-Wheeler gauge are, for $\ell\geq 1$,
\begin{align}
H^{(0)}_{\ell 0} ={}&  H^{(1)}_{\ell0} = - H^{(2)}_{\ell0}= -f_{\ell0}\, e^{-t/2m}\, \delta(U),
\nonumber\\
K_{\ell m}={}& 0, \qquad f_{\ell0}=4\sqrt{\pi}{\cal N}_H p_H \frac{\sqrt{2\ell+1}}{\ell(\ell+1)+1}  . \label{HK}
\end{align}
Only the $m=0$ spherical harmonics are excited, since our particle is at the north pole. The  metric perturbations to the  Schwarzschild black hole caused by the plunging of a particle close to the horizon  never leave the horizon. In fact, the metric perturbations, which are produced by the plunging particle, are localized by the $\delta(U)$ factor in Eq. \eqref{HK}, on the future horizon. Since the latter has no support away from $U=0$, the perturbations cannot be extended to the exterior of the horizon, and thus there are no modes that can be detected by a distant observer.

This can also be seen from  the gauge invariant Zerilli-Moncrief master function $\Psi_{\text{ZM}}=\sum_{\ell,m} \Psi_{\text{ZM}}^{\ell m}Y_{\ell m}(\Omega)$, where $\Psi_{\text{ZM}}^{\ell m}$ are given by
 \cite{Martel:2005ir} 
\begin{align}
\Psi_{\text{ZM}}^{\ell 0} &= 
\frac{8m\, f_{\ell0}}{\ell (\ell+1)\Lambda} \left(\frac{r}{2m}-1\right) e^{-t/2m}\,  \delta(U), \label{U}
\end{align}
where 
\begin{eqnarray}
\Lambda=(\ell-1)(\ell+2)+\frac{6m}{r}.
\end{eqnarray}
Clearly then, since $r=2m$ at $U=0$, we get  
\begin{eqnarray}
\Psi^{\ell0}_{\rm ZM}=0,
\end{eqnarray} 
which is another way to see that there are no propagating  RWZ  degrees of freedom. However, the vanishing of the master function does not imply  that the Dray-'t Hooft geometry is trivial. Its physical content is in the shift $\Delta V$ of the null coordinate $V$ at horizon crossing 
\begin{equation}
\Delta V=V_+-V_-=f(\Omega),
\end{equation}
where $V_+$ and $V_-$ are the values of $V$ on the two sides of the horizon.

As a final note, let us also notice a nice connection to causality. 
Since $V=e^{\kappa v}$, a geodesic crossing the shock at advanced time $v_-$ acquires a Shapiro-like delay 
\begin{equation}
\Delta v=\frac1\kappa\ln\bigl(1+f\,e^{-\kappa v_-}\bigr)
=\frac1\kappa\sum_{n\geq1}\frac{(-1)^{n+1}}{n}\,f^{n}\,e^{-n\kappa v_-}.
\label{eq:DTladder}
\end{equation}
This expansion involves the advanced coordinate $v$ and it is the advanced time counterpart of the near horizon exponential structure. It should not be identified with the outgoing ladder in the  retarded coordinate $u$. We can show that $\Delta v$ is always positive as required by causality arguments \cite{Camanho:2014apa}. 

Notice also that this construction may extend to other nonextremal stationary horizons when an appropriate shock solution and horizon constraint exist.

\medskip\noindent
\textbf{
\hspace{-0.2cm}\
\emph{Conclusions.}}
We have identified the physical origin of the purely damped sequence
$\omega_n=-in\kappa$ that supplements quasinormal ringing in the plunge
signal, and argued that it never reaches a distant observer. Two results
support this claim.

First, for any field analytic in the exterior Kruskal coordinate $U$, the
ladder is nothing but the Taylor expansion of that field re-read through
$U=-e^{-\kappa u}$: its frequencies are dilation weights of the
near-horizon boost, equivalently a lattice of Mellin poles at the horizon
endpoint, fixed by regularity and surface gravity alone. These poles
belong to the local near-horizon representation, and 
their local appearance does not by itself imply poles of the global retarded Green function. This is consistent with, and explains the frequency spacing found in, the global result of Ref.~\cite{Kuntz:2026xep}, which shows that the complete source-convolved waveform of a plunging particle contains no isolated redshift terms.

Second, we have shown where the analytic premise fails: in the late-time regime in which the near-horizon source must be resummed into the shock. Taking a family of timelike worldlines to the horizon in a controlled ${\cal E}\to0$ limit with $m_{\rm P}/{\cal E}$ fixed, we found that the point-particle stress tensor converges, not to a smooth source, whose field would be regular in $U$ and admit the Taylor expansion above, but distributionally to a null source confined to $U=0$:  precisely the source of the Dray-'t~Hooft shock. The resulting metric is an exact, nonperturbative solution of Einstein's equations, smooth on either side of the horizon but genuinely singular across it, with no analytic expansion in $U$ and hence no
finite, regulator-independent set of Taylor or Mellin coefficients.

Together these results clarify the status of the horizon modes. As a
spectrum of Mellin poles, the ladder is a legitimate but purely local
artifact of near-horizon regularity, not an emission mechanism. 
As the response to the plunge, sourced by a shock rather than an analytic
field, it does not exist even at the local level: its would-be Taylor
coefficients are ill-defined in the zero-width limit. The horizon modes
should therefore be understood as a singular completion of the analytic
near-horizon sector, not as an additional tower of radiative degrees of
freedom; they do not leave the horizon.

\medskip
\noindent
{\itshape Acknowledgments.} D.G.  acknowledges support from  the National Science and Technology Council (NSTC) of Taiwan with the Young Scholar Columbus Fellowship grant 114-2636-M-110-004 and 115-2112-M-110-010.   A.R.  acknowledges support from the Swiss National Science Foundation (project number CRSII5\_213497). 

\bibliographystyle{JHEP}
\bibliography{refsv2f}

\end{document}